\documentclass[preprint,showpacs,preprintnumbers,nofootinbib,amsmath,amssymb,aps,prl,longbibliography]{revtex4-1}
\usepackage{graphicx}
\usepackage{dcolumn}
\usepackage{bm}
\begin{document}
\title{Canonical and microcanonical ensemble descriptions \\ of thermal pairing
within BCS and quasiparticle RPA}
\author{N. Quang Hung$^{1}$}
 \email{nqhung@riken.jp}
 \altaffiliation[On leave of absence from the ]
 {Center for Nuclear Physics, Institute of Physics, Hanoi, Vietnam}
 \author{N. Dinh Dang$^{1,2}$}
  \email{dang@riken.jp}
\affiliation{1) Theoretical Nuclear Physics Laboratory, RIKEN
Nishina Center for Accelerator-Based Science,
2-1 Hirosawa, Wako City, 351-0198 Saitama, Japan\\
2) Institute for Nuclear Science and Technique, Hanoi, Vietnam}
\date{\today}
\begin{abstract}
    We propose a description of pairing properties in finite
    systems within the canonical and microcanonical
    ensembles. The approach is derived by solving the BCS and self-consistent
    quasiparticle random-phase approximation with the Lipkin-Nogami
    particle-number projection at zero temperature. The obtained eigenvalues
    are embedded into the canonical and microcanonical
    ensembles. The results obtained are found in quite good agreement with the exact solutions of the
    doubly-folded equidistant multilevel pairing model as well as the experimental data
    for $^{56}$Fe nucleus. The merit of the present
    approach resides in its simplicity and its application to a wider
    range of particle number, where the exact solution is impracticable.
\end{abstract}

\maketitle

The study of thermodynamic properties of finite systems such as
atomic nuclei or ultra-small metallic grains has been an
important topic in nuclear physics. Thermodynamic properties
are usually described within the grand canonical ensemble (GCE),
canonical ensemble (CE), or microcanonical ensemble (MCE). The GCE consists of identical
systems in thermal equilibrium, each of which exchanges its energy
and particle number with the external reservoir. The systems of the
CE exchange only their energies whereas the particle number is
fixed. The MCE consists of
thermally isolated systems with the same energy and particle
number. Among these ensembles, the GCE is often used in the
theoretical studies because of convenience in the calculations of
thermodynamic quantities. For example, the well-known BCS theory of
superconductivity \cite{BCS} was derived based on the GCE. This
theory describes very well thermodynamic properties of infinite systems such
as metal superconductors, where quantal and thermal fluctuations
are zero. The latter, however, have been shown to be quite
significant in finite small systems \cite{Moretto,SPA,Zele,MBCS,FTBCS1,Ensemble}.
They smoothed out the superfluid-normal (SN) phase transition, which is a
typical property of infinite systems.
Most of theoretical approaches to thermal pairing
have been derived so far within the GCE in finite
systems such as atomic nuclei, where no particle-number fluctuations are allowed.
Therefore, the CE and MCE should be used instead. Moreover,
the exact eigenvalues of the pairing problem~\cite{Exact}
are usually embedded into the CE and MCE~\cite{Ensemble, Sumaryada}.
But this task is impracticable for particle numbers $N>$ 14
in the case of half-filled doubly-folded multilevel model with
$N=\Omega$ ($\Omega$ is number of single-particle levels). The reason is
that the exact partition function at finite temperature $T\neq$ 0
should include all excited states.
The finite-temperature shell-model Monte Carlo (FTSMMC) \cite{SMMC1, SMMC2}
has also been used to evaluate the partition function, but it is very time consuming
and cannot be applied to heavy nuclei.
The FTSMMC cannot be applied to the MCE either because it is
impossible to include all the
microstates of the system. It is therefore highly desirable
to construct an approach based on the CE and MCE,
which can offer results in good agreement with the
exact solutions for any value of the particle number,
as well as with experimental data for realistic nuclei.
This is the goal of the present study.

We consider the pairing Hamiltonian $H=\sum_{j}\epsilon_j
\hat{N}_j-G\sum_{jj'}\hat{P}_{j}^{\dagger}\hat{P}_{j'}~$,
where $\hat{N}_{j}=\sum_{m}a_{jm}^{\dagger}a_{jm}$ is the
particle-number operator, and
$\hat{P}_{j}=\sum_{m>0}a_{jm}^{\dagger}a_{j\tilde{m}}^{\dagger},
\hat{P}_j=(\hat{P}_j^{\dagger})^{\dagger}$ are the pairing
operators. The operators $a_{jm}^{\dagger}$ and $a_{jm}$ are respectively
the creation and destruction operators of a nucleon moving on the $j$-th orbital
($a_{j\tilde{m}}=(-)^{j-m}a_{j-m}$). This Hamiltonian has been
diagonalized exactly in  \cite{Exact} to obtain n$_{\rm
Exact}$=$\sum_{S}{\rm C}_{S}^{\Omega}\times {\rm C}_{N_{\rm
pair}-S/2}^{\Omega-S}$ eigenstates with eigenvalues
$\varepsilon_{S}^{\rm Exact}$ and occupation numbers $f_{j}^{S}$,
where S = 0, 2, ... $\Omega$ is the total seniority of the system and C$_{m}^{n}=m!/[n!(m-n)!]$
\cite{Ensemble}. The exact partition function is constructed by
embedding the exact eigenvalues into the CE as
$Z_{\rm Exact}(\beta)=\sum_{S}d_{S}\exp({-\beta\varepsilon_{S}^{\rm
Exact}})$~, with the degeneracy $d_{S}=2^{S}$ and inverse temperature
$\beta=1/T$. Knowing the partition function $Z$, one calculates
the free energy $F$, entropy $S$, total energy ${\cal E}$,
heat capacity $C$,  and pairing gap $\Delta$ as $F = -T{\rm ln}Z(T)$,
$S = -{\partial F}/{\partial T}$, ${\cal E} = F + TS$,
$C ={\partial{\cal E}}/{\partial T}$, and $\Delta=[-G({\cal
E}-2\sum_{j}\epsilon_{j}f_{j}+G\sum_{j}f_{j}^{2})]^{1/2}$,
where $f_{j}$ are the occupation numbers on the $j$th orbital
obtained by averaging the state-dependent occupation numbers $f_j^{(S)}$
within the CE~\cite{Ensemble}.
The conventional BCS theory at $T\neq$ 0 (FTBCS) is derived
by using the variational procedure with respect to the quasiparticle
Hamiltonian ${\cal H}$. The latter is obtained by applying the
Bogoliubov's transformation of the pairing Hamiltonian $H$ from the
particle operators, $a_{jm}^{\dagger}$ and $a_{jm}$, to the
quasiparticle ones, $\alpha_{jm}^{\dagger}$ and $\alpha_{jm}$. The
variational procedure is carried out on the average quantities within
the GCE, namely $\langle{\cal\hat{O}}\rangle \equiv {\rm
Tr}[{\cal\hat{O}}e^{-\beta({\cal H}-\lambda\hat{N})}] / {\rm
Tr}[e^{-\beta({\cal H}-\lambda\hat{N})}]$ for any observable $\hat{O}$
with $\lambda$ being the
chemical potential determined as the Lagrangian multiplier to preserve
the average particle number $\langle\hat{N}\rangle=N$.
Therefore the conventional FTBCS theory is called the
GCE-BCS in the present article.

The CE description of thermal pairing can be
undertaken in two directions. The first one is to apply the exact
particle-number projection (PNP) at finite temperature on top of the GCE
theory~\cite{Nakada, PNP}. This approach is
rather complicated for realistic nuclei. The second direction, which
the present study follows, is to embed the solutions of a
theoretical approximation at zero temperature, which conserves the
particle number, into the CE. We employ the solutions of the BCS
equations with PNP within the Lipkin-Nogami
method \cite{LN} (LNBCS) for each total seniority $S$ and embed the
eigenvalues $\varepsilon_{S}^{\rm LNBCS}$ obtained at $T$=0 into the CE.
The partition function of the so-called CE-LNBCS approach is then
given as
\begin{equation}
Z_{\rm LNBCS}(\beta)=\sum_{S}d_{S}e^{-\beta\varepsilon_{S}^{\rm LNBCS}}
~. \label{ZBCS}
\end{equation}
Using it, we can obtain the free energy,
entropy, total energy and heat capacity as has been discussed above,
by replacing $Z_{\rm Exact}(\beta)$
with $Z_{\rm LNBCS}(\beta)$. As for the pairing gap, it is obtained
by averaging the gaps $\Delta_{S}^{\rm LNBCS}$, which are the solutions of the
LNBCS equations at $T=0$, namely
\begin{equation}
\Delta={Z_{\rm LNBCS}(\beta)^{-1}}\sum_{S}\Delta_{S}^{\rm
LNBCS}d_{S}e^{-\beta\varepsilon_{S}^{\rm LNBCS}}~. \label{DeltaBCS}
\end{equation}
Within the BCS theory at $T=0$, only the lowest
eigenstates can be obtained, e.g the ground-state energy for $S$=0.
The total number of the LNBCS eigenstates embedded into the CE is
equal to n$_{\rm LNBCS}=\sum_{S}{\rm C}_{S}^{\Omega}$, which is much
smaller than n$_{\rm Exact}$. The CE of these lowest eigenstates is
therefore comparable with the exact one only in the region of low
$T$. At higher $T$, one needs to include not only the ground state but
also excited states into the CE.
This can be resolved by means of
the selfconsistent quasiparticle RPA
with Lipkin-Nogami PNP (LNSCQRPA)~\cite{SCQRPA}.
The LNSCQRPA includes the ground-state and screening
correlations, which are neglected within the conventional BCS and QRPA.
The importance of these correlations in finite systems
has been demonstrated in  \cite{SCQRPA}. They improve the agreement between the
energies of ground state and low-lying excited states obtained
within the LNSCQRPA and the corresponding exact results for
the doubly-folded multilevel pairing model. The formalism of the LNSCQRPA was
presented in details in  \cite{SCQRPA}, so we do not repeat it
here. The total number of eigenstates obtained within the LNSCQRPA
is n$_{\rm LNSCQRPA}=\sum_{S}{\rm C}_{S}^{\Omega}\times(\Omega-S) >
$n$_{\rm BCS}$ because of the presence of QRPA excited
states~\footnote{The first solution of the SCQRPA equations corresponds to the
spurious mode and is subtracted from the total
number of solutions.}, but it is still much smaller than n$_{\rm
Exact}$. For example, for the half-filled model with $\Omega=N=10$,
n$_{\rm Exact}=8953$, whereas n$_{\rm LNBCS}=512$ and n$_{\rm
LNSCQRPA}=2561$. The thermodynamic quantities are obtained within
the CE-LNSCQRPA in the same way as that for the CE-LNBCS (\ref{ZBCS}), namely from
the CE-LNSCQRPA partition function $Z_{\rm
LNSCQRPA}(\beta)=\sum_{S}d_{S}\exp[-\beta\varepsilon_{S}^{\rm
LNSCQRPA}]$, where $\varepsilon_{S}^{\rm LNSCQRPA}$ are the eigenvalues obtained
by solving the LNSCQRPA equations for each total seniority $S$.

Within the MCE description, we use the eigenvalues $\varepsilon_{S}^{\rm LNBCS}$
and $\varepsilon_{S}^{\rm LNSCQRPA}$
to calculate the MCE entropy directly from the Boltzmann's definition
$S({\cal E})={\rm ln}{W}({\cal E})$, where ${W}({\cal E})$ is the number of
accessible states within the energy interval $({\cal E},{\cal E}+\delta{\cal
E})$ \cite{Ensemble, EXP}. The corresponding approaches,
which embed the LNBCS and LNSCQRPA eigenvalues at $T=0$ into the MCE, are called
the MCE-LNBCS and MCE-LNSCQRPA, respectively.
\begin{figure}
     \includegraphics[width=8.7cm]{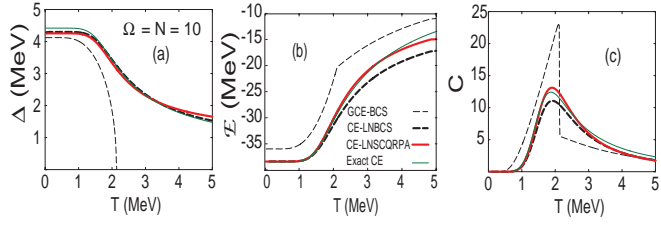}
     \caption{(Color online) Pairing gap $\Delta$,
     total energy ${\cal E}$ and heat capacity $C$ as functions of $T$
     obtained within the Richardson model for $N=$10 with $G$=1 MeV.
     The thin dashed, thick dashed, thick solid, and thin solid lines denote
     the GCE-BCS, CE-LNBCS, CE-LNSCQRPA and exact CE results, respectively ~.
     \label{N10}}
\end{figure}
\begin{figure}
     \includegraphics[width=7.5cm]{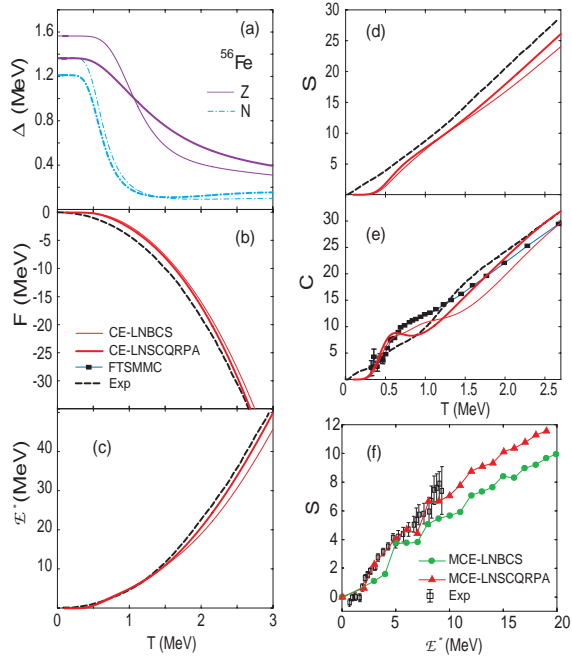}
     \caption{(Color online) Pairing gap $\Delta$, free energy $F$,
     excitation energy ${\cal E}^{*}$, heat capacity $C$,
     CE and MCE entropies $S$ as functions of $T$ for $^{56}$Fe.
     In (a), the solid and dash-dotted lines denote the neutron and proton
     gaps, respectively, with the thin and thick lines  corresponding
     to CE-LNBCS and CE-LNSCQRPA results, respectively.
     In (b) -- (e), the thin and solid lines denote the CE-LNBCS and CE-LNSCQRPA
     results, whereas the dashed line and the full squares connected by solid line stand for the
     experimental data and the FTSMMC results, respectively.
     In (f), the MCE entropies obtained within the MCE-LNBCS (circles),
     MCE-LNSCQRPA (triangles) are plotted vs experimental data (open squares
     with error bars)~.
     \label{Fe56}}
\end{figure}

The proposed approach is tested within a schematic model and applied
to describe thermodynamic quantities for
a realistic nucleus. For the schematic model, we employ the
so-called Richardson (or ladder) model, which consists of $\Omega$ doubly-folded equidistant levels with the
single-particle energies chosen as
$\epsilon_{j}= j - (\Omega + 1)/2$ MeV.
The model is half-filled, i.e. $N=\Omega$. As for the
realistic nucleus, we employ the axially deformed Woods-Saxon single-particle
spectra for Fe$^{56}$ nucleus with the depth of the central potential
$V=V_0[1\pm k(N-Z)/(N+Z)]$, where $V_{0}=$ 51.0 MeV, $k=$ 0.86,
the diffuseness $a$=0.67 fm, $r_0$=1.27 fm \cite{WS}, the deformation parameter
$\beta_2=$ 0.244~\cite{SMMC1}, and the spin-orbit strength $\lambda=$
35.0.

Shown in Fig. \ref{N10} are the pairing gaps $\Delta$, total
energies ${\cal E}$ and heat capacities $C$ obtained
within the GCE-BCS, CE-LNBCS, CE-LNSCQRPA as well as the exact CE for
the Richardson model with $N = $10 and $G$ = 1 MeV. The figure clearly
shows that the CE-LNSCQRPA results (thick solid lines) nearly coincide with the
exact CE (thin solid lines) for all thermodynamic quantities
under consideration. A slight difference between the CE-LNSCQRPA
results and the exact ones at very high $T$ occurs because
the CE-LNSCQRPA includes only the low-lying excited
states in the ensemble. These states are known to be important in
low and medium temperature regions. At very high $T$, one needs to
include high-lying excited states into the ensemble as well.
The same feature is seen from the results obtained
for other systems with different $N$ and $G$. The results obtained
within the CE-LNBCS (thick dashed lines) are a bit off from the
exact ones but, as compared to the predictions by the GCE-BCS (thin
dashed lines), they still offer much better agreement with the exact
solutions. The most interesting feature is that neither the
pairing gaps obtained within the CE-LNBCS nor those obtained within the CE-LNSCQRPA
collapse at the critical temperature $T_C$ as predicted by the
GCE-BCS, but they all monotonously decrease with increasing $T$,
just like the exact CE gap. Consequently, the sharp peak in the heat
capacity, which is the signature of SN phase transition, is smoothed
out within these approaches. This feature implies that the CE-LNBCS
and CE-LNSCQRPA include the effects of quantal and thermal fluctuations, which
are neglected in the GCE-BCS. Moreover, it also suggests
that, if any exact PNP method at finite $T$
could bring the GCE-BCS to the CE-BCS, the results obtained within
this projected CE-BCS should have the same qualitative behavior as
the CE-LNBCS results presented in present article with a nonvanishing
gap.

The pairing gap, free energy, excitation energy, heat capacity and
entropy in $^{56}$Fe nucleus are displayed in Fig. \ref{Fe56}. In
order to have a consistent comparison with the recent experimental
data in  \cite{EXP} as well as the FTSMMC results for these
quantities, we employ the same configurations proposed in
\cite{SMMC2}, namely the CE-LNBCS and CE-LNSCQRPA are calculated
within the complete $pf+0g_{9/2}$ major shell. The results obtained
within this shell are then combined with those obtained within the
independent-particle model (IPM) by using Eq. (15) of
 \cite{SMMC2}, where $Z^{'}_{\nu,tr}$ in the present case
is the CE-LNBCS or CE-LNSCQRPA partition function, whereas
$Z^{'}_{sp}$ and $Z^{'}_{sp,tr}$ are the CE partition functions
obtained within the IPM for full (from bottom to $h_{11/2}$ orbital)
and truncated (from $f_{7/2}$ to $h_{11/2}$ orbitals) model spaces,
respectively (The width of resonance single-particle states is
neglected for simplicity). The parameter $G$ is adjusted so that the
pairing gaps for protons (Z) and neutrons (N) obtained within the
LNBCS at $T=$ 0 reproduce the experimental values, namely
$\Delta_{Z}\simeq $1.57 MeV and $\Delta_{N}\simeq$ 1.36 MeV
\cite{EXP}. It is clear from this figure that the CE-LNSCQRPA
results agree quite well with the experimental data, which are also
deducted from the CE. The small difference between the CE-LNSCQRPA
results and experimental data at high $T$ is probably due to the
fact that latter have been extracted from the experimental level
density rather than being obtained from direct measurements. The
level density, however, cannot be measured at high excitation energy
$E^{*}>$ 10 MeV, but has been instead obtained from the back-shifted
Fermi gas (BSFG) formula (see e.g. Eq. (2) of  \cite{EXP}), assuming
a zero pairing gap in this region of excitation energy. As we can
see now, this approximation is rather crude because the pairing gap
is always finite even at high $E^{*}>$ 10 MeV (or $T>T_{C}$) because
of thermal fluctuations. This is demonstrated in Fig. \ref{Fe56}
(a), which clearly shows that, at $T >$ 1.5 MeV, which corresponds
to $E^{*} >$ 10 MeV, the neutron and proton gaps are still finite.
This feature puts the use of BSFG formula to extrapolate the
measured level density to high $E^{*}$ under question. As regards
the heat capacity obtained from the experimental data plus BSFG
[Fig. \ref{Fe56} (e)], it only shows an oscillation rather than a
peak or even an S shape in the temperature region around $T_{C}$,
whereas all CE-LNBCS, CE-LNSCQRPA as well as FTSMMC results predict
a bump in this region. The absence of a bump at $T\sim T_{C}$ in the
experimentally extracted heat capacity for $^{56}$Fe also seems to
contradict those obtained for nuclei in rare-earth region by using
the same method, where a clear S shape or even a broad bump has been
seen~\cite{Kaneko}. Regarding the difference shown in Fig.
\ref{Fe56} (e) between the FTSMMC results and ours at high $T$, it
might come from the absence of quadrupole interaction in our
calculations. It has been well-known that the multipole
interactions, in particular the quadrupole force, which is important
for quadrupole deformation, can have significant effects on the
nuclear level density at moderate and high temperatures (See Fig. 8
and Table 1 in Ref. ~\cite{DangZP}). Fluctuations of other degrees
of freedom such as nuclear shapes due to rotation may also play an
important role. Most probably these effects will show up at high
temperatures and angular momenta, which are considered neither in
the experimental data for $^{56}$Fe that we use nor within the
present framework of our model.

A genuine thermodynamic observable is the MCE entropy at low $E^{*}$ because it is calculated
directly from the measured level density (See e.g. Eq. (3) of  \cite{EXP}).
Shown in Fig. \ref{Fe56} (f) are the MCE entropies obtained within the MCE-LNBCS and MCE-LNSCQRPA
using the Boltzmann's definition with
$\delta{\cal E}$ = 1 MeV versus the experimental data.
It is seen that the MCE-LNSCQRPA entropy not only
offers the best fit to the experimental data
but also predicts the results up to higher $E^{*} >$ 10 MeV.

In summary, we have proposed two approximations, which embed the
solutions of the BCS and SCQRPA with Lipkin-Nogami PNP at $T=0$ into
the CE and MCE. The proposed approaches are tested within the
Richardson model and applied to describe the recent experimentally
extracted thermodynamic quantities for $^{56}$Fe nucleus. The
analysis of numerical calculations for the pairing gap, free energy,
total energy, heat capacity and entropy shows that the CE-LNSCQRPA
predictions are in quite good agreements with the exact results
(whenever the latter are available), as well as the recent
experimental data. Moreover, this is a microscopic approach that can
describe simultaneously and selfconistently all the experimentally
extracted quantities, namely the free energy, total energy, heat
capacity, and entropy within both CE and MCE treatments. To our
knowledge, this is the first time that the recent experimental MCE
entropy~\cite{EXP} has been successfully described by a consistent
microscopic theory. It also shows that the SN phase transition
predicted by the conventional GCE-BCS theory is smoothed out even
within the CE-LNBCS calculations due to the effects of quantal and
thermal fluctuations, leading a nonvanishing pairing gap. The
results of the present study put under question the use of BSFG
formula to extrapolate the measured level density to high excitation
energy. The merit of the present approach resides in its simplicity
and its feasibility in the application to larger finite systems,
where the exact matrix diagonalization and/or solving the Richardson
equation are impracticable to find all eigenvalues, whereas the
FTSMMC method is time consuming.

The authors thank Peter Schuck (Orsay) for fruitful
discussions. NQH is a Nishina Memorial Fellow at RIKEN.
The numerical calculations were carried out using the FORTRAN IMSL
Library by Visual Numerics
on the RIKEN Integrated Cluster of Clusters (RICC) system.


\begin{thebibliography}{99}
\bibitem{BCS}
J. Bardeen, L. Cooper, and Schrieffer, Phys. Rev. \textbf{108}, 1175
(1957).
\bibitem{Moretto}
L.G. Moretto, Phys. Lett. B \textbf{40}, 1 (1972);
A.L. Goodman, Phys. Rev. C \textbf{29}, 1887 (1984);
J.L. Egido, P. Ring, S. Iwasaki, and H.J. Mang, Phys. Lett. B {\bf
154}, 1 (1985).
\bibitem{SPA}
R. Rossignoli, P. Ring and N.D. Dang, Phys. Lett. B \textbf{297}, 9
(1992); N.D. Dang, P. Ring and R. Rossignoli, Phys. Rev. C
\textbf{47}, 606 (1993).
\bibitem{Zele}V. Zelevinsky, B.A. Brown, N. Frazier, and M. Horoi,
Phys. Rep. {\bf 276}, 85 (1996).
\bibitem{MBCS}
N. Dinh Dang and V. Zelevinsky, Phys. Rev. C {\bf 64}, 064319
(2001); N. Dinh Dang and A. Arima, Phys. Rev. C {\bf 67}, 014304
(2003); N.D. Dang and A. Arima, Phys. Rev. C {\bf 68}, 014318
(2003); N.D. Dang, Nucl. Phys. A {\bf 784}, 147 (2007).
\bibitem{FTBCS1}
N. Dinh Dang and N. Quang Hung, Phys. Rev. C {\bf 77}, 064315
(2008); N.Q. Hung and N.D. Dang, Phys. Rev. C {\bf 78}, 064315
(2008).
\bibitem{Ensemble}
N.Q. Hung and N.D. Dang, Phys. Rev. C {\bf 79}, 054328 (2009).
\bibitem{Exact}
R.W. Richardson, Phys. Lett. {\bf 3}, 277 (1963); Ibid. {\bf 14},
325 (1965); A. Volya, B.A. Brown, and V. Zelevinsky, Phys.
Lett. B {\bf 509} (2001) 37.
\bibitem{Sumaryada}
T. Sumaryada and A. Volya, Phys. Rev. C {\bf 76}, 024319 (2007).
\bibitem{SMMC1}
S. Liu and Y. Alhassid, Phys. Rev. Lett {\bf 87}, 022501 (2001).
\bibitem{SMMC2}
Y. Alhassid, G. F. Bertsch, and L. Fang, Phys. Rev. C {\bf 68}, 044322 (2003).
\bibitem{Nakada}
K. Tanabe and H. Nakada, Phys. Rev. C {\bf 71}, 024314 (2005); H.
Nakada and K. Tanabe, Phys. Rev. C {\bf 74}, 061301(R) (2006).
\bibitem{PNP}
R. Rossignoli and P. Ring, Ann. Phys. (NY) 235, 350 (1994), R.
Rossignoli, P. Ring, and N. D. Dang, Phys. Lett. B297, 9 (1992).
\bibitem{LN}
H. J. Lipkin, Ann. Phys. (NY) \textbf{9} 272 (1960); Y. Nogami, Phys.
Lett. \textbf{15} 4 (1965).
\bibitem{SCQRPA}
N.Q. Hung and N.D. Dang, Phys. Rev. C {\bf 76}, 054302 (2007);
Ibid. {\bf 77}, 029905(E) (2008).
\bibitem{WS}
S. Cwiok {\it et al.}, Comput. Phys. Commun. {\bf 46}, 379 (1987).
\bibitem{EXP}
E. Algin {\it et al.}, Phys. Rev. C {\bf 78}, 054321 (2008).
\bibitem{Kaneko} A. Schiller {\it et al.}, Phys. Atom. Nucl. {\bf 64}, 1186
(2001); K. Kaneko {\it et al.}, Phys. Rev. C {\bf 74}, 024325
(2006).
\bibitem{DangZP}
N. Dinh Dang, Z. Phys. A {\bf 335}, 253 (1990).
\end{thebibliography}
\end{document}